\documentclass[preprint,12pt]{elsarticle}

\journal{Spatial Statistics }

\usepackage{amsmath}
\usepackage{enumitem}
\usepackage{graphicx} 
\usepackage{longtable, tabu}
\usepackage{booktabs}
\usepackage{caption}
\graphicspath{ {./img/} }
\usepackage{natbib}
\usepackage[utf8]{inputenc}
\usepackage[T1]{fontenc}
\usepackage{geometry}

\newcommand\iid{i.i.d.}
\newcommand\pN{\mathcal{N}}

\begin{document}

\begin{frontmatter}

\title{Feasible pairwise pseudo-likelihood inference on spatial regressions in irregular lattice grids: the KD-T PL algorithm}

\author[1]{Giuseppe Arbia}
\ead{giuseppearbia13@gmail.com}
\author[1]{Niccolò Salvini\corref{cor1}}%
\ead{niccolo.salvini27@gmail.com}
\cortext[cor1]{Corresponding author}
\affiliation[1]{organization={Università Cattolica del Sacro Cuore, Department of Statistical Sciences},
addressline={L.go Gemelli 8},
postcode={00168},
city={Rome},
country={Italy}}

\begin{abstract}

Spatial regression models are central to the field of spatial statistics. Nevertheless, their estimation in case of large and irregular gridded spatial datasets presents considerable computational challenges. To tackle these computational problems, Arbia \citep{arbia_2014_pairwise} introduced a pseudo-likelihood approach (called pairwise likelihood, say PL) which required the identification of pairs of observations that are internally correlated, but mutually conditionally uncorrelated. However, while the PL estimators enjoy optimal theoretical properties, their practical implementation when dealing with data observed on irregular grids suffers from dramatic computational issues (connected with the identification of the pairs of observations) that, in most empirical cases, negatively counter-balance its advantages. In this paper we introduce an algorithm specifically designed to streamline the computation of the PL in large and irregularly gridded spatial datasets, dramatically simplifying the estimation phase. In particular, we focus on the estimation of Spatial Error models (SEM). Our proposed approach, efficiently pairs spatial couples exploiting the KD tree data structure and exploits it to derive the closed-form expressions for fast parameter approximation. To showcase the efficiency of our method, we provide an illustrative example using simulated data, demonstrating the computational advantages if compared to a full likelihood inference are not at the expenses of accuracy.
\end{abstract}


\end{frontmatter}

\section{Introduction}
Spatial regression models are fundamental tools in the domain of spatial statistics, enabling researchers to explore and analyse the relationship between spatially dependent data, which  arise in various scientific disciplines and real-world applications. These models offer insights into the spatial relationships that exist between observations, allowing for the characterisation of spatial patterns and the investigation of the underlying processes. However, their application to large and irregularly gridded spatial datasets poses significant computational challenges, which have long been a bottleneck in the analysis (\citep{arbia_2014_a}, \citep{bivand_2018_comparing}, \cite{banerjee_2015_hierarchical}, \citep{banerjee_2015_hierarchical}, \citep{elhorst_2010_applied} ).

In an effort to address the computational complexities inherent in spatial regression on irregular grids, \citep{arbia_2014_pairwise} introduced a promising approach called "pairwise likelihood" (PL) which belongs to the class of composite likelihood (\citep{varin_2011_an}, \citep{varin_2005_a}). The PL approach hinges on identifying pairs of observations that exhibit internal spatial correlation while being mutually conditionally uncorrelated. Although the PL approach possesses appealing theoretically properties, its practical implementation often gives rise to formidable computational hurdles. In many empirical scenarios, the advantages of PL estimators are overshadowed, particularly when confronted with large dataset observed on irregular grids, by the computational difficulties they introduce.

The present paper seeks to introduce a novel algorithm tailored to expedite the PL computation in the context of large and irregularly gridded spatial datasets, thereby significantly simplifying the estimation phase. Our primary focus revolves around the estimation of Spatial Error models (SEM), a class of spatial regression models widely used in the spatial statistics literature ( \citep{lesage_2009_introduction} \citep{arbia_2014_a}, \citep{anselin_2002_under}).

Our proposed algorithm, termed the "KD-Tree Pairwise Likelihood" (hereafter KD-T PL), leverages the power of the KD Tree data structure \citep{bentley_1975_multidimensional} to efficiently pair spatial observations, contributing to a streamlined estimation process, but also maintaining stochastic spatial Independence among paired couples. The algorithm also capitalises on closed-form expressions derived from the paired data, which enable swift parameter approximation. To underscore the efficiency of our method, we present the results of a Monte Carlo experiment through which we showcase the computational advantages and the heightened accuracy achieved with the KD-T approach in comparison to traditional full likelihood inference methods.

This paper not only addresses the pivotal issue of computational efficiency in spatial regression, but also illustrates the substantial benefits of the KD-T algorithm for handling large and irregularly gridded spatial datasets. By offering an innovative solution to the computational challenges associated with spatial regression, our work advances the toolkit available to researchers and practitioners working with large spatial data, facilitating efficient analyses of complex spatial irregular relationships. In the following sections \ref{comput-iss-spat-regr} to \ref{sim-study-and-comp}, we will delve into the details of the KD-T PL algorithm, its implementation, and its performance in practical applications, providing a comprehensive resource for spatial statisticians seeking to make the most of their irregular medium-large grid spatial data. Section \ref{concl} concludes.

\section{Computational Issues in spatial regression} \label{comput-iss-spat-regr}
Spatial regression models are widely used in the field of spatial statistics to analyse the relationship between a dependent variable and one or more independent variables in a geospatial context (\citep{arbia_2014_a}, \cite{cressie_1992_statistics}). These models provide valuable insights into spatial patterns and can be used to make predictions for unobserved locations \citep{haining_2000_spatial}.
One commonly used spatial regression model is the Spatial Error Model (SEM) \cite{arbia_2014_a} which assumes that the errors in the regression model are spatially autocorrelated. The SEM model can be expressed as follows: 

\begin{equation}\label{eq:model-1}
y = X \beta + u
\end{equation}

\begin{equation}\label{eq:model-2}
u = \rho W + \varepsilon
\end{equation}

where $\varepsilon$ is $\iid~ \sim \pN(0, \sigma)$ and W is an n-by-n weight matrix describing the topology of the spatial system. In the structural alternative formulation, model \ref{eq:model-1} and \ref{eq:model-2} can be written as:

\begin{equation}\label{eq:topology-model}
 y = X \beta + (I_n - \rho W)^{-1} \varepsilon 
\end{equation}

The task of estimating the SEM parameters when dealing with large and irregularly gridded spatial datasets can be extremely challenging for a number of reasons. First of all, the definition of the weight matrix ($W$) can be be memory-intensive and time-consuming, especially for large datasets with complex spatial structures forcing $W$ to be large and sparse. A second major problem concerns the fact that SEM models require a large number of matrix multiplications involving $W$, a task that can be computationally expensive, especially when dealing with large and dense $W$ matrices, the computational complexity directly depending on matrix size and multiplication algorithms. Naive methods scale with the order of $O(n^3)$, though modern libraries and software for numerical computations often use highly optimised matrix multiplication algorithms that take advantage of hardware features like parallel processing (utilising multiple CPU cores or GPUs) and cache memory to speed up the computations. A third crucial element that creates computational problems is represented by the estimation method. Indeed, the estimation of the model parameters often involves iterative numerical techniques as it happens, e. g., when using Maximum  Likelihood (ML) or Instrumental Variables (IV) estimation. \citep{lesage_2009_introduction}.
These methods may require multiple iterations, making the estimation process time-consuming, particularly for large datasets. For a spatial error model, the likelihood function is based on the assumption that the errors ($\epsilon$) follow a multivariate normal distribution. The likelihood function considers both the traditional error term and the spatially correlated error term. To find the ML estimation of the parameters $\beta$ and $\rho$ in Equation (3), optimisation techniques, such as the Newton-Raphson method or gradient descent are often employed. The alternative estimation method based on on the Feasible GLS (FGLS) \citep{kelejian_1997_estimation}, does not require the hypothesis of normality and reduces, but does not fully eliminate the computational problems \citep{ghiringhelli_2022_recursive}.

As a matter of fact the relevance of computational issues in spatial regression analysis as a matter of fact is a longstanding challenge, as it is witnessed by the large number of approximate solutions suggested in the literature over the years, see \citep{ghiringhelli_2022_recursive} for a review.

Many popular alternatives to the full likelihood approach intervenes to correct one or more of the problems mentioned above; these include the GMM approach \citep{kelejian_1997_estimation}, partial likelihood \citep{cox_1975_partial}, the m-th order likelihood \citep{azzalini_1983_maximum}, the matrix exponential spatial specification, \citep{Pace2004}. Additional efforts have also been devoted to embed approximations techniques with software and hardware tweaks, such as Parallel and Distributed Computing \citep{gerber_2021_parallel}.

\section{A review of the pairwise likelihood inference }

The pairwise likelihood (PL) inference method is an estimation method falling in the category of pseud-likelihood family, introduced by \citep{arbia_2014_pairwise} to facilitate the estimation of spatial econometric models. Unlike  most of the spatial econometric methods that require the prior specification of a weight matrix, PL inference doesn't needs such assumptions; instead, it exploits the inherent spatial relationships within the dataset.
Indeed, the PL approach represents a generalisation of the so-called "coding technique" proposed in \citep{besag_1974_spatial}. The original Besag's proposal suggested to select single observations that are far enough in space to be assumed spatially independent, and to derive a pseudo-likelihood as the product of the associated univariate density functions. Proceeding in this way, the method does not allow the possibility of estimating spatial dependence effects. In contrast, the PL method  \citep{varin_2005_a}, \citep{arbia_2014_pairwise} suggests to select pairs of neighbouring observations that are internally correlated, but that are far enough to be considered independent on each other. In this way we can preserve the relevant spatial information that was lost in the original "coding technique". 
Formally, consider a spatial dataset constituted by $n$ locations, and let us select a subset of, say, $q$ locations $(l = 1, ..., q)$ . Let us also randomly choose a second set of $q$ locations $(l = 1, ..., q)$  within the neighbourhood of each location $i$, say $N\left(i\right)$. We can assume that the two random variables constituting each pair (say $\epsilon_i$ and $\epsilon_l$), are spatially dependent due to their proximity, but we can also assume that the pairs $\{ \epsilon_i, \epsilon_l \}$ and $\{ \epsilon_j, \epsilon_k \}$ are stochastically independent if are able to identify the pairs in such a way that both $j$ and $k$ do not belong to the joint neighborhood of $i$ and $l$, say $N\left(i, l\right)$.

Given the previous definitions, the PL can be defined as the product of the bivariate likelihoods for all pairs:

\begin{equation}\label{eq:likelihood-pairs}
L_p(\theta)=\prod_{(i, l)} L_{i l}(\theta)     
\end{equation}

$L_{i l}(\theta)$ denoting the bivariate joint probability distribution of locations $i$ and $l$.
This approach does not require a full specification of the matrix  $W$, dramatically reduces the computational burden and enables interesting insights into the interpretation of parameter estimates. It also allows the mathematical derivation of a closed forms solution for the parameters' estimators, the derivation of the Fisher information matrix and of the estimators' asymptotic properties (see \cite{arbia_2014_pairwise} for details). 

In particular for a SEM model \citep{arbia_2014_a} we can assume that any pair of errors in Equation (2) follows a joint bivariate Gaussian distribution such that:

\begin{equation}\label{eq:gauss-distr}
\left(\begin{array}{l} \varepsilon_i 
\varepsilon_l \end{array}\right) \approx B V N\left(0_2 I_2, \sigma_l^2 \Omega\right) 
\end{equation}

With $I_2$ a 2-by-2 identity matrix, the correlation matrix $\Omega$ is expressed by:

\begin{equation}\label{eq:corr-mat}
\Omega=\left(\begin{array}{cc} 1 & \psi \\
\psi & 1 \end{array}\right) 
\end{equation}

with $\psi \in(-1 ;+1)$. The bivariate density function can then be written as:

\begin{equation}\label{eq:biv-den-funct}
f_{\varepsilon_i \varepsilon_l}\left(\varepsilon_i \varepsilon_l\right)=\frac{1}{2 \pi \sigma^2 \sqrt{1-\psi^2}} \exp \left\{-\frac{1}{2 \sigma^2\left(1-\psi^2\right)}\left[\varepsilon_i^2-2 \psi \varepsilon_i \varepsilon_l+\varepsilon_l^2\right]\right\} \quad \text { if } l \in N(i) 
\end{equation}

leading to the following log-likelihood expression:

\begin{equation}\label{eq:full-lik-expre}
L(\theta)=L\left(\beta, \sigma^2, \psi\right)=\prod_{i=1}^q f_{\varepsilon_i \varepsilon_l}\left(\varepsilon_i \varepsilon_l\right) 
\end{equation}

We can then introduce the following six sufficient statistics with obvious notation:


\begin{align}\label{eq:suff-stat}
\alpha_1 &= \sum_{i=1}^q x_i^2+\sum_{l=1}^q x_l^2=\sum_{j=1}^{2 q} x_j^2, & \alpha_2 &= \sum_{i=1}^q y_i^2+\sum_{l=1}^q y_l^2=\sum_{j=1}^{2 q} y_j^2 \\
\alpha_3 &= \sum_{i=1}^q x_i y_i+\sum_{l=1}^q x_l y_l=\sum_{j=1}^{2 q} x_j y_j, & \alpha_4 &= \sum_{i=1}^q x_i y_l+\sum_{l=1}^q x_l y_i \\
\alpha_5 &= \sum_{i=1}^q x_i x_l, & \alpha_6 &= \sum_{i=1}^q y_i y_l .
\end{align}

Finally, the PL estimators $\hat{\beta}_{PL}$,$\hat{\sigma}_{PL}$,  $\hat{\psi}_{PL}$  can be obtained as the solution of the following system of equations:

\begin{equation}\label{eq:beta-hat}
\hat{\beta}_{PL} =\frac{\alpha_3-\hat{\psi}_{PL} \alpha_4}{\alpha_1-2 \hat{\psi}_{PL} \alpha_5}   
\end{equation}


\begin{equation}\label{eq:sigma-hat}
\hat{\sigma}_{PL}^2 = \frac{\alpha_2+\hat{\beta}_{PL}^2 \alpha_1-2 \hat{\beta}_{PL} \alpha_3-2 \hat{\psi}_{PL} \alpha_6-2 \hat{\psi}_{PL} \hat{\beta}_{PL}^2 \alpha_5+2 \hat{\psi}_{PL} \hat{\beta}_{PL} \alpha_4}{2 q\left(1-\hat{\psi}_{PL}^2\right)}
\end{equation}


\begin{equation}\label{eq:psi-hat}
\hat{\psi}_{PL}=\frac{\alpha_6-\hat{\beta}_{PL} \alpha_4+\hat{\beta}_{PL}^2 \alpha_5}{q \hat{\sigma}_{PL}^2}    
\end{equation}

\section{The KD-T PL coupling algorithm}

In section 3 we presented the PL approach to SEM regression model estimation. However, from a practical point of view, the process of selecting neighbouring pairs that are internally correlated and independent to each other can be challenging due to the fact that the distance matrix scales quadratically with $n$. This situation may sterilise computational gain derived from closed from expressions reported in Equations \ref{eq:beta-hat}, \ref{eq:sigma-hat} and \ref{eq:psi-hat}. To mitigate this adverse effect in the present paper we propose a novel approach that incorporates the KD-tree (short for k-dimensional tree) data structure into the coupling process.
The KD tree, is a powerful data structure used in various computational applications, particularly in multidimensional data organisation and search \citep{bentley_1975_multidimensional}. Its primary purpose is to efficiently partition and organise data points in such a way that it can facilitate rapid nearest neighbour searches.

\begin{figure}
    \includegraphics[width=\textwidth]{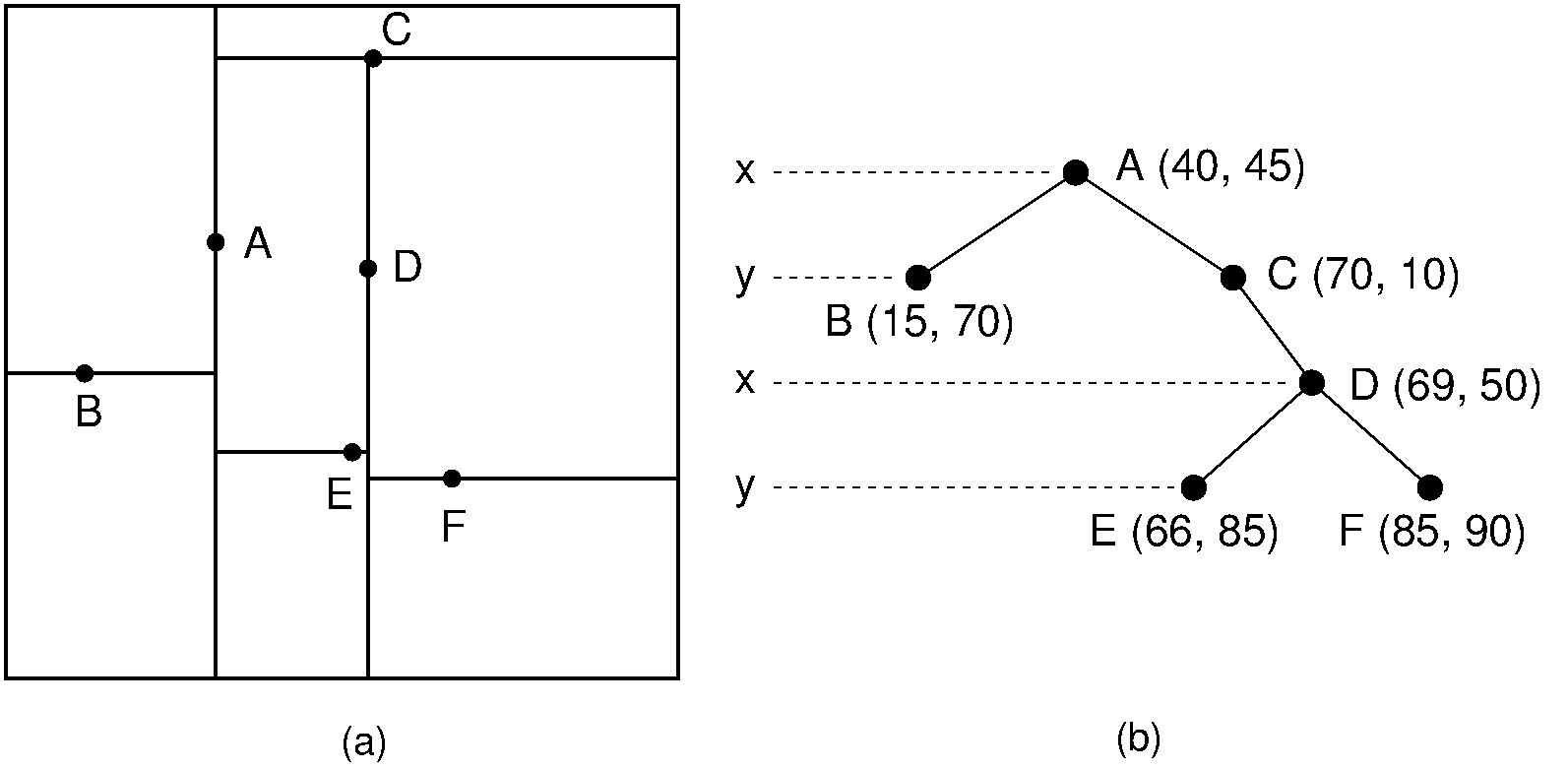}
    \caption{Example of a KD tree. (a) The KD tree decomposition for a 128×128 - unit region containing seven data points. (b) The KD tree for the region. }
    \label{fig:kdtree}
    \centering
\end{figure}

Consider a dataset with $n$  data points, each residing in a $k$-dimensional space, where $k$  represents the number of attributes or features. The KD tree algorithm recursively divides this multidimensional space into axis-aligned hyperplanes, essentially creating a binary tree structure. At each level of the tree, the algorithm selects one dimension to partition the data, alternating between dimensions at each level.
The partitioning process proceeds as follows:
\begin{itemize}
    \item \textbf{STEP 1:} Choose a dimension $d$ (e.g. $d = 1,2,3 .. k$ ) based on some criterion, often the dimension with the maximum variance.
    \item \textbf{STEP 2: } Find the median value of the data points along dimension $d$, This median value serves as the splitting threshold.
    \item \textbf{STEP 3:} Divide the data points into two subsets: those with values less than the median along dimension $d$ (left subtree) and those with values greater than or equal to the median along dimension $d$ (right subtree). Figure \ref{fig:kdtree}  (a) shows how the space is divided into regions, with each region containing one data point.
\end{itemize}

The divisions are made by creating axis-aligned splits at each level of the tree. These splits are determined by the data points that would act as nodes if we were to visualise the tree in three dimensions. In this 2-dimensional example, the splits are along the x or y axis. The algorithm alternates between axes at each level, a procedure that helps to balancing the tree while ensuring efficient searches.

In Figure \ref{fig:kdtree} (b) we report a graphical representation of the binary tree created by the KD-tree algorithm with respect to the region subdivisions made still in Figure \ref{fig:kdtree} (a). The points from A through F are arranged in such a manner that reflects their division in space. In this binary tree structure, each node represents a data point, and a "leaf" node (that is the end point of a branch) contains the final subset of points. For our specific use, we are interested in creating couples, which means that each leaf node will contain exactly two data points.

The KD-tree assists in rapidly identifying these couples by recursively dividing the space and sorting points such that each division step brings spatially closer points together. For instance, if we are at a leaf node representing a vertical split, we know that the two points in this node are much closer to each other along the x-axis than to any points in adjacent leaf nodes. This principle applies to every level of the tree and across both dimensions.

Moreover when we perform a search for the nearest neighbours to form couples, the KD-tree allows us to quickly exclude large portions of the search space because we can eliminate entire regions that do not fall within the desired radius. Therefore, instead of comparing every point with every other point (and operation that would be computationally intractable for large data sets), we limit our search to the points contained within the same or adjacent leaf nodes, which greatly reduces the computational burden. This approach aligns perfectly with the need to select spatially dependent pairs while ensuring that the pairs are also spatially independent from each other in the dataset.

As a matter of fact the primary goal of this algorithm is to create couples of data points while also ensuring spatial independence between them. 

\begin{figure}
    \includegraphics[width=\textwidth]{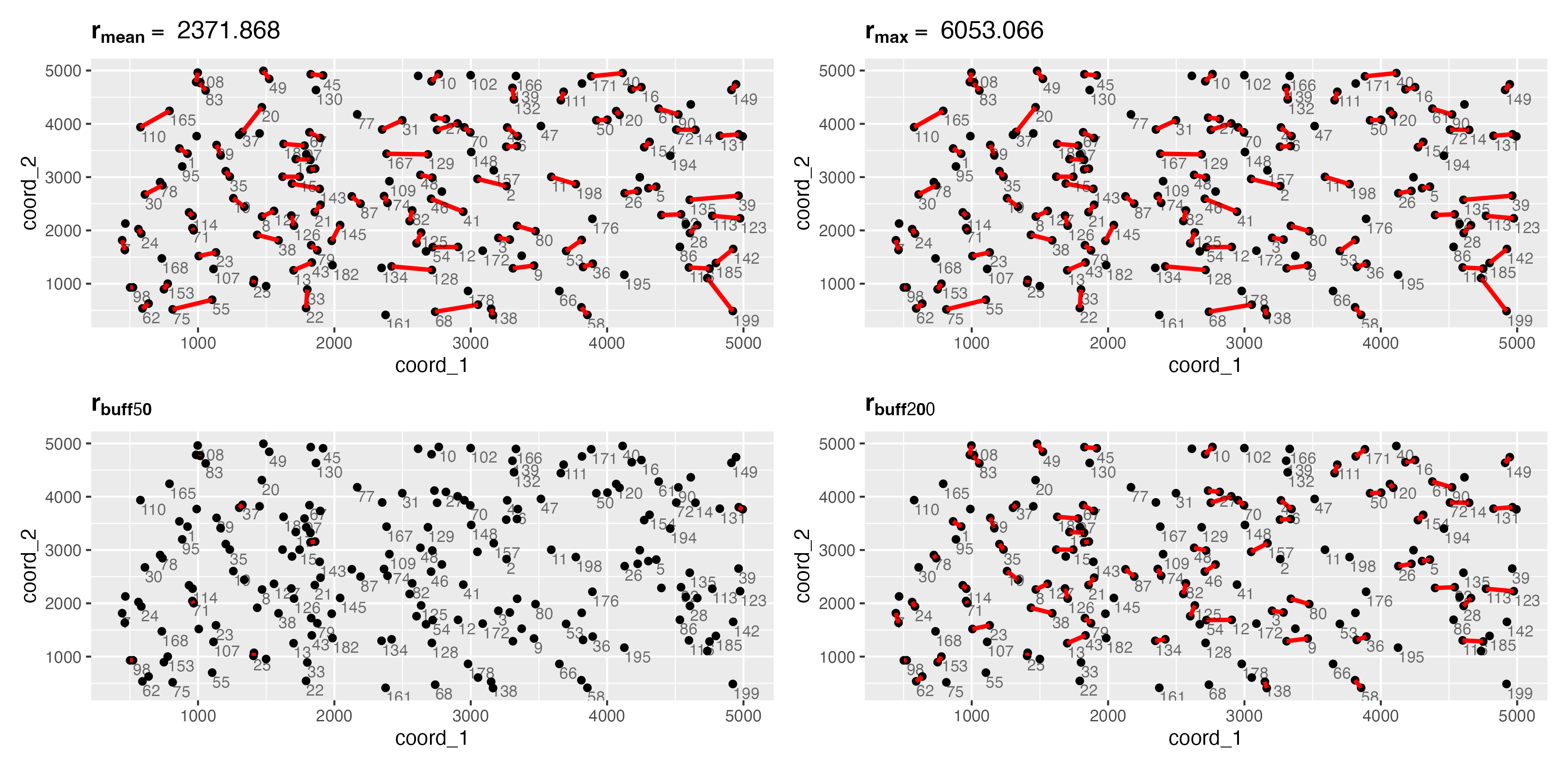}
    \caption{Visualisation of spatial point pairings using pairwise KD-Tree at varying radius parameters. The top row illustrates dense (left) and sparse (right) neighbour connections with mean and maximum radius values, respectively. The bottom row displays the effect of reduced radius buffers on neighbour identification, with minimal connections at $r_{buff50}$ (left) and slightly increased pairings at $r_{buff200}$ (right), demonstrating the algorithm's adaptability to different spatial independence criteria in large, irregularly gridded datasets.}
    \label{fig:buff_couples}
    \centering
\end{figure}

In fact we utilise  KD-tree to facilitate nearest neighbour search, which has a time complexity of approximately $O(nlog(n))$, but we also we take advantage of the KD-tree's radius search, which allows us to find neighbours within a specified radius. This parameter is essential for controlling the level of neighbourhood (within a specified radius), while preserving stochastic spatial independence between couples. The default radius is set as the mean distance between each data points and is denoted as $r_{\text{mean}}$. Figure \ref{fig:buff_couples}  shows the outcome of 4 different search procedures using 4 different $r$ radius. 

The algorithm begins by computing the nearest neighbours for each data point with respect to a chosen radius. In order to efficiently find the two nearest neighbours for each data point within the radius $r$, we utilise the KD-tree algorithm, making use of the RANN R package, which is a binder for C++ Approximate Nearest Neighbour Searching (ANN) KD-tree implementation \citep{arya_2019_rann}. 

Formally let $P$ represent the set of data points, and for each $p \in P$, we aim to find two other points, $p_1$ and $p_2$, such that: $p_1, p_2 \in P $ but $p_1$ and $p_2$ are not equal to $p$, and the distance from $p$ to $p_1$, denoted as $d(p, p_1)$ is less than or equal the distance from $p$ and $p_2$ ($d(p, p_2)$), within a radius $r$. Here, $d(p, q)$ represents the Euclidean distance between two points $p$ and $q$ but different distance definitions can be adopted to suit any specific problem.

To facilitate the computation after, we initialise an empty $n \times n$ matrix denoted as $D$ which will store the pairwise distances between data points within the specified radius $r$. We then create two empty lists, say $C_{couplets}$ and $C_{paired}$ to store and keep track of points that have already been paired.

Our algorithm initially pairs up data points such that each point is only paired once. Let $C_{paired}$ represent the set of points that have been already paired. For each point $p \in P$ , we find two points $p_1$ and $p_2$ such that $p_1 \comma, p_2 \in P$ and $d(p, p_1) \leq d(p, p_2)$ within the radius $r$.  This pairing is achieved by exploiting the nearest neighbours KD-tree indexes. We store the couplet in the 
$C_{couplets}$ and update $C_{paired}$ accordingly. After forming a couplet $(p_1, p_2)$ we remove the paired points $p_1$ and $p_2$ from the remaining set of points. Simultaneously, we update the distance matrix $D$  with the pairwise distances for the coupled points: $D_{i,j} = D_{j,i} = d(p_1, p_2)$. The algorithm proceed further until all points are paired.

\section{Simulation study and comparisons} \label{sim-study-and-comp}

Using simulated data, we show that our approach is much faster than the full likelihood approach and provides accurate estimates of the model's parameters. We adopt the same simulation scheme described in \cite{arbia_2014_pairwise}, and we generate synthetic spatial data following these steps:

\begin{enumerate}
  \item We constructed a correlation matrix $\Phi$  using the Inverse Exponential Correlation function i. e. 
        
        \begin{equation}\label{eq:inv-corr-funct}
            \Phi(\mathbf{d}, \phi)=e^{-\phi \cdot \mathbf{d}}
        \end{equation}
  where, $\mathbf{d}$ represents the scaled distance matrix, and $\phi$  is a distance decay parameter. To ensure a valid correlation matrix, we make the necessary adjustments by setting the matrix diagonal elements to 1.

  \item We apply the Cholesky decomposition to $\Phi$ in order to obtain a lower triangular matrix $\mathbf{L}$. This step ensures that the generated spatial data preserves the specified spatial correlation structure.
  
  \item We then simulate the random effect vector $z$. This is achieved by drawing samples from a standard normal distribution with mean 0 and standard deviation 1. Specifically, the vector $z$ is generated as follows:
    \begin{equation}\label{eq:random-effect}
        z \sim \mathcal{N}(0,1)
    \end{equation}

  \item We then generate $\epsilon$ by multiplying the lower triangular matrix $\mathbf{L}$ obtained from the Cholesky decomposition of $\Phi$ with the random effect vector $z$. This can be represented as:
  
    \begin{equation}\label{eq:epsilon}
        \epsilon = \mathbf{L} \cdot z
    \end{equation}

  \item Finally, with the spatial data matrix $\mathbf{X}$ and introducing random errors $\epsilon$ produced in Equation \ref{eq:epsilon} we simulate the dependent variable $Y$.
  
    \begin{equation}\label{eq:dependent-variable}
        Y = \boldsymbol{\beta} \cdot \mathbf{X} + \boldsymbol{\epsilon}
    \end{equation}
    where $\boldsymbol{\beta}$ represents the regression coefficients set equal to 1 for simplicity. Indeed this assumption can be streched and the model can be extended to multivariate regression \citep{arbia_2014_pairwise}.

\end{enumerate}

To allow comparison, we estimate the model's parameters $\beta$, $\sigma$ and $\psi$ using (i) the pairwise likelihood (PL) described in Section 3 and (ii) the full Maximum Likelihood (FL) procedure \footnote{In particular we used the spatialreg R package to estimate Full Likelihood \citep{bivand_2013_applied}, \citep{bivand_2015_comparing}}, exploiting default numerical optimisation \citep{bivand_2015_comparing} and accounting for the entire data likelihood. To comprehensively assess method performance, we vary the correlation parameter $\phi$ (equal to 0.8 or 1) and the sample size $n$ (assuming the values $n$ of  200, 800, 1800, 5000). The number of Monte Carlo replication is set equal to 100. We conduct a Monte Carlo experiment by iterating through various combinations of $\psi$ and $n$.
For each parameter combination, the following steps are executed:

\begin{enumerate}[label=(\roman*)]
    \item generate random spatial data with specified  $\psi$ and $n$
    \item estimate spatial regression parameters $\phi$ using PL and FL methods.
    \item for a generic paramater $\theta$ calculate the bias $\mathbf{B}$ , expressed as:
        \begin{equation}\label{eq:bias-expr}
            \mathbf{B}_{PL}=\left|\theta-\operatorname{Ave}\left(\hat{\theta}_{PL, k}\right)\right|
        \end{equation}
        the relative bias $\mathbf{RB}$, expressed as:
        \begin{equation}\label{eq:rel-bia-expr}
            \mathbf{R B}_{PL}=\frac{\left|\theta-\operatorname{Ave}\left(\hat{\theta}_{PL, k}\right)\right|}{|\theta|}
        \end{equation}
        and the mean squared error $\mathbf{M S E}$, expressed as:
        \begin{equation}\label{eq:mse-expr}
            \mathbf{M S E}=\mathbf{A v e}\left[\hat{\theta}_{PL, k}- \mathbf{A v e}\left(\hat{\theta}_{PL, k}\right)\right]^2+B_{PL}^2
        \end{equation}
        where $\mathbf{A v e}( . )$ is the average across iterations for each parameter estimate.
    \item  Store the results systematically for analysis. Specific enphasis needs to be devoted to the parameter estimates $\hat{\beta}_{PL}$ and $\hat{\beta}_{FL}$. Some other statistics are also reported so as to validate simulation goodness and model compliance to the data generation process assumptions. As a matter of fact, by design, the data generating process does not favour any specific estimation methods.
        
\end{enumerate}

\setlength{\LTpost}{0mm}
\begin{longtable}{c c c c c c c}
\caption{
Monte Carlo simulation results for coefficients comparing Pairwise Likelihood (PL) exploiting KD-T PL algorithm and the full Maximum Likelihood (FL) methods for spatial regression parameter estimation. The table presents estimates for standard deviation ($\sigma$), spatial autocorrelation ($\psi$), and regression coefficient ($\beta$) across different sample sizes ($n$) and distance decay parameters ($\phi = 1$ strong spatial correlation; $\phi = .8$ weak spatial autocorrelation).
}\label{tab:mc-sim-coeffs} \\ 
\toprule
 & \multicolumn{2}{c}{$\beta = 1$} & \multicolumn{2}{c}{$\sigma = 1$} & \multicolumn{2}{c}{$\psi$} \\ 
\cmidrule(lr){2-3} \cmidrule(lr){4-5} \cmidrule(lr){6-7}
$n$ & $\hat{\beta}_{PL}$ & $\hat{\beta}_{FL}$ & $\hat{\sigma}_{PL}$ & $\hat{\sigma}_{FL}$ & $\hat{\psi}_{PL}$ & $\hat{\rho}_{FL}\textsuperscript{\textit{1}}$ \\ 
\midrule\addlinespace[2.5pt]
\multicolumn{7}{l}{$\phi= 1$} \\ 
\midrule\addlinespace[2.5pt]
200 & 0.99517 & 1.01481 & 0.99898 & 1.07092 & 0.33860 & 0.00148 \\ 
800 & 1.00203 & 1.01172 & 1.00401 & 1.09072 & 0.33977 & 0.00041 \\ 
1800 & 1.00101 & 0.99803 & 1.00452 & 0.96011 & 0.34780 & -0.00046 \\ 
5000 & 1.00106 & 0.99998 & 1.01006 & 0.99515 & 0.33602 & -0.00003 \\ 
\midrule\addlinespace[2.5pt]
\multicolumn{7}{l}{ $\phi = 0.8$} \\ 
\midrule\addlinespace[2.5pt]
200 & 0.99539 & 1.01481 & 0.99194 & 1.07092 & 0.27317 & 0.00148 \\ 
800 & 1.00211 & 1.01172 & 0.99756 & 1.09072 & 0.27383 & 0.00041 \\ 
1800 & 1.00107 & 0.99803 & 0.99787 & 0.96011 & 0.28241 & -0.00046 \\ 
5000 & 1.00109 & 0.99998 & 1.00300 & 0.99515 & 0.27120 & -0.00003 \\ 
\bottomrule
\end{longtable}
\begin{minipage}{\linewidth}
\textsuperscript{\textit{1}}We still report $\hat{\rho}_{FL}$, see Equation \ref{eq:model-2} for completeness purposes even though it can not be compared with $\hat{\psi}_{PL}$\\
\end{minipage}

\newpage

\setlength{\LTpost}{0mm}
\begin{longtable}{c c c c c c c}
\caption{
The relative bias ($\mathbf{R B}$) and mean squared error ($\mathbf{M S E}$) indicate the precision and accuracy of the KD-T PL approach, showcasing its effectiveness in efficiently producing accurate estimators for large spatial datasets in the presence of varying degrees of spatial correlation
}\label{tab:mc-sim-met}  \\ 
\toprule
 & \multicolumn{3}{c}{$\mathbf{R B} (\textbf{PL}, \textbf{FL})$} & \multicolumn{3}{c}{ $\mathbf{M S E} (\textbf{PL}, \textbf{FL})$} \\ 
\cmidrule(lr){2-4} \cmidrule(lr){5-7}
$n$ & $\hat{\beta}$ & $\hat{\sigma}$ & $\hat{\psi}$ & $\hat{\beta}$ & $\hat{\sigma}$ & $\hat{\psi}$  \\ 
\midrule\addlinespace[2.5pt]
\multicolumn{7}{l}{$\phi= 1$} \\ 
\midrule\addlinespace[2.5pt]
200 & $\begin{array}{c} (0.00483, \\ 0.01481) \end{array}$ & $\begin{array}{c} (0.07092, \\ 0.07092) \end{array}$ & $\begin{array}{c} (0.66140, \\ 0.99852) \end{array}$ & $\begin{array}{c} (0.00002, \\ 0.00022) \end{array}$ & $\begin{array}{c} (0.00000, \\ 0.00503) \end{array}$ & $\begin{array}{c} (0.43745, \\ 0.99703) \end{array}$ \\ 
800 & $\begin{array}{c} (0.00203, \\ 0.01172) \end{array}$ & $\begin{array}{c} (0.09072, \\ 0.09072) \end{array}$ & $\begin{array}{c} (0.66023, \\ 0.99959) \end{array}$ & $\begin{array}{c} (0.00000, \\ 0.00014) \end{array}$ & $\begin{array}{c} (0.00002, \\ 0.00823) \end{array}$ & $\begin{array}{c} (0.43590, \\ 0.99919) \end{array}$ \\ 
1800 & $\begin{array}{c} (0.00101, \\ 0.00197) \end{array}$ & $\begin{array}{c} (0.03989, \\ 0.03989) \end{array}$ & $\begin{array}{c} (0.65220, \\ 1.00046) \end{array}$ & $\begin{array}{c} (0.00000, \\ 0.00000) \end{array}$ & $\begin{array}{c} (0.00002, \\ 0.00159) \end{array}$ & $\begin{array}{c} (0.42536, \\ 1.00092) \end{array}$ \\ 
5000 & $\begin{array}{c} (0.00106, \\ 0.00002) \end{array}$ & $\begin{array}{c} (0.00485, \\ 0.00485) \end{array}$ & $\begin{array}{c} (0.66398, \\ 1.00003) \end{array}$ & $\begin{array}{c} (0.00000, \\ 0.00000) \end{array}$ & $\begin{array}{c} (0.00010, \\ 0.00002) \end{array}$ & $\begin{array}{c} (0.44088, \\ 1.00005) \end{array}$ \\ 
\midrule\addlinespace[2.5pt]
\multicolumn{7}{l}{$\phi= 0.8$} \\ 
\midrule\addlinespace[2.5pt]
200 & $\begin{array}{c} (0.00461, \\ 0.01481) \end{array}$ & $\begin{array}{c} (0.07092, \\ 0.07092) \end{array}$ & $\begin{array}{c} (0.90854, \\ 1.24815) \end{array}$ & $\begin{array}{c} (0.00002, \\ 0.00022) \end{array}$ & $\begin{array}{c} (0.00006, \\ 0.00503) \end{array}$ & $\begin{array}{c} (0.52829, \\ 0.99703) \end{array}$ \\ 
800 & $\begin{array}{c} (0.00211, \\ 0.01172) \end{array}$ & $\begin{array}{c} (0.09072, \\ 0.09072) \end{array}$ & $\begin{array}{c} (0.90772, \\ 1.24949) \end{array}$ & $\begin{array}{c} (0.00000, \\ 0.00014) \end{array}$ & $\begin{array}{c} (0.00001, \\ 0.00823) \end{array}$ & $\begin{array}{c} (0.52733, \\ 0.99919) \end{array}$ \\ 
1800 & $\begin{array}{c} (0.00107, \\ 0.00197) \end{array}$ & $\begin{array}{c} (0.03989, \\ 0.03989) \end{array}$ & $\begin{array}{c} (0.89699, \\ 1.25057) \end{array}$ & $\begin{array}{c} (0.00000, \\ 0.00000) \end{array}$ & $\begin{array}{c} (0.00000, \\ 0.00159) \end{array}$ & $\begin{array}{c} (0.51494, \\ 1.00092) \end{array}$ \\ 
5000 & $\begin{array}{c} (0.00109, \\ 0.00002) \end{array}$ & $\begin{array}{c} (0.00485, \\ 0.00485) \end{array}$ & $\begin{array}{c} (0.91100, \\ 1.25003) \end{array}$ & $\begin{array}{c} (0.00000, \\ 0.00000) \end{array}$ & $\begin{array}{c} (0.00001, \\ 0.00002) \end{array}$ & $\begin{array}{c} (0.53114, \\ 1.00005) \end{array}$ \\ 
\bottomrule
\end{longtable}


\hspace{1cm} 

Table \ref{tab:mc-sim-met} shows the relative bias and the mean squared error obtained using the PL and the FL estimation method. It suggests that both methods provide point estimates that are close to the true values. Furthermore, the relative bias and the MSE of the PL approach is consistently low and comparable with the one obtained with the FL approach, thus underscoring the accuracy of the PL estimates. This implies that the advantages of PL estimation does not compromise the quality of parameter estimates which are shown in Table \ref{tab:mc-sim-coeffs}. 

\begin{figure}
    \includegraphics[width=\textwidth]{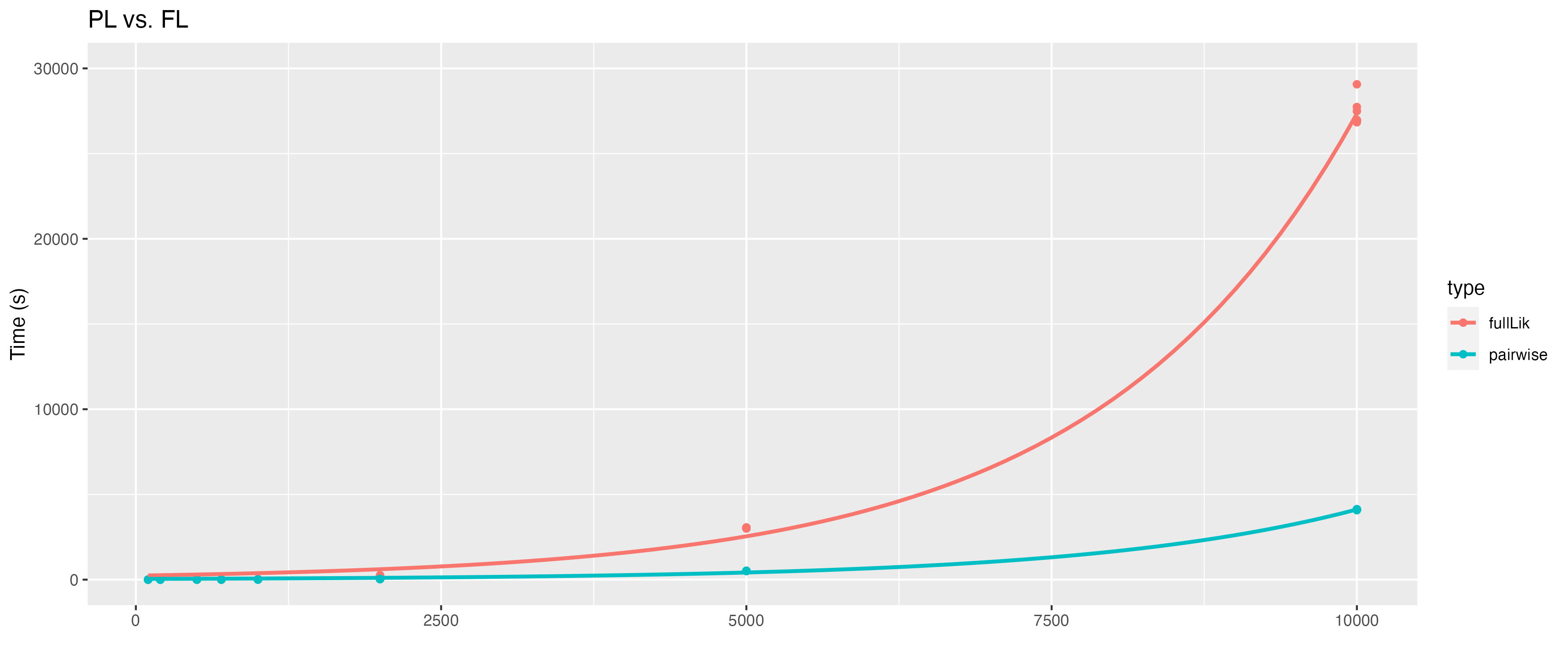}
    \caption{Performance comparison of Pairwise Likelihood (PL) using the KD-Tree Pairwise Likelihood (KD-T PL) algorithm versus Full Likelihood (FL) method over increasing sample sizes.}
    \label{fig:comput_times}
    \centering
\end{figure}

Most notably, Figure \ref{fig:comput_times}, shows the computational time involved with the two estimation methods as a function of the sample size. The visual inspection of the graph shows that the KD-T PL algorithm demonstrates substantial computational advantages if compared to the ML method. Indeed, not only the time required by the coupled KD-T PL is reduced with respect to the full ML approach, but also it is only marginally affected by the sample size while with the ML the computation time increases exponentially with $n$. This makes our proposed approach particularly appealing for applications involving large spatial datasets, where time constraints are a critical concern.

We repeated the Monte Carlo experiment described above to explore also the effect of varying the radius within the KD-T PL algorithm. As we have already mentioned, the radius $r$, plays a crucial role in defining the neighbourhood structure for the spatial observations. By modifying the radius, we aim to investigate its impact on the performance of the Pairwise Likelihood (PL) estimation method. In this experiment, we altered the radius $r$, ranging from small values ($r$=50) to larger ones ($r$=800), thus representing the spatial scale of neighbourhood relationships. Specifically, we considered the value of $r_{mean}$ (the mean distance across all distances),  and $r_{max}$, (the max pairwise observed distance). We then evaluated various buffer radii values: 50, 200, 350, 500, 650, and 800 units, denoted respectively as $r_{buff50}$, $r_{buff200}$, $r_{buff350}$, $r_{buff500}$, $r_{buff650}$, and $r_{buff800}$. Each radius value, $r_{buff}$, represents the additional circular distance buffer added to the mean radius $r_{mean}$, symbolised as $r_{buffh}$ where $h$ is the buffer distance.

Tables \ref{tab:mc-sim-coefs-buff} and \ref{tab:mc-sim-met-buff} provide a detailed analysis of how the buffer radius impacts the estimation of spatial regression parameters.

When comparing results in Tables \ref{tab:mc-sim-coefs-buff} and \ref{tab:mc-sim-met-buff} from Tables \ref{tab:mc-sim-coeffs} and \ref{tab:mc-sim-met} without buffer adjustments, we can see that as the buffer increases, there are slight variations in the parameter estimates for both PL and FL methods. The relative bias and $\mathbf{MSE}$ for $\sigma$ and $\psi$ seem to remain relatively stable across buffer changes, indicating that the KD-T PL algorithm is robust to variations in the neighbourhood definition.

Notably, the buffer's influence on $\beta$'s estimation appears minimal, suggesting that the regression coefficient estimates are not heavily dependent on the neighbourhood radius within the tested range. This is a positive indication that the KD-T PL algorithm can be applied flexibly across different spatial scales without significant loss of accuracy.

However, the buffer radius does affect the PL method's efficiency. Larger buffers tend to increase computational time due to the expanded search space for neighbours. The results indicate that an optimal buffer size exists where the balance between accuracy and computational efficiency is achieved. The analysis underscores the importance of carefully choosing the radius to ensure optimal performance of the KD-T PL algorithm in different spatial data contexts.

\setlength{\LTpost}{0mm}
\begin{longtable}{cccccccc}
\caption{
Monte Carlo simulation coefficients results depicting the impact of varying buffer distances on parameter estimation accuracy for Pairwise Likelihood (PL) using the KD-T PL algorithm versus Full Likelihood (FL) methods. The table presents estimations for $\sigma$, $\psi$, and $\beta$ under different buffer scenarios.
}\label{tab:mc-sim-coefs-buff} \\ 
\toprule
 &  & \multicolumn{2}{c}{$\beta = 1$} & \multicolumn{2}{c}{$\sigma = 1$} & \multicolumn{2}{c}{$\psi$} \\ 
\cmidrule(lr){3-4} \cmidrule(lr){5-6} \cmidrule(lr){7-8}
buffer & $n$ & $\hat{\beta}_{PL}$ & $\hat{\beta}_{FL}$ & $\hat{\sigma}_{PL}$ & $\hat{\sigma}_{FL}$ & $\hat{\psi}_{PL}$ & $\hat{\rho}_{FL}\textsuperscript{\textit{1}}$ \\ 
\midrule\addlinespace[2.5pt]
\multicolumn{8}{l}{ $\phi= 1$} \\ 
\midrule\addlinespace[2.5pt]
50 & 200 & 1.06990 & 1.03728 & 0.98808 & 0.98784 & 0.21365 & -0.00550 \\ 
50 & 800 & 0.99812 & 1.00970 & 1.03009 & 0.96549 & 0.39517 & -0.00004 \\ 
50 & 1800 & 1.00633 & 0.99269 & 1.00046 & 1.02971 & 0.36517 & -0.00009 \\ 
50 & 5000 & 1.00238 & 0.99892 & 0.99390 & 0.98174 & 0.32672 & -0.00007 \\ 
200 & 200 & 0.99461 & 1.03434 & 0.92551 & 0.97336 & 0.32841 & -0.00423 \\ 
200 & 800 & 0.99325 & 1.01254 & 0.98491 & 0.96251 & 0.39313 & 0.00010 \\ 
200 & 1800 & 1.00419 & 0.99383 & 1.01309 & 1.02065 & 0.37842 & -0.00009 \\ 
200 & 5000 & 1.00176 & 0.99898 & 0.98630 & 0.97756 & 0.33294 & -0.00007 \\ 
350 & 200 & 1.00347 & 1.03784 & 0.94073 & 0.97138 & 0.22902 & -0.00480 \\ 
350 & 800 & 0.99254 & 1.01238 & 0.98620 & 0.96630 & 0.39177 & 0.00009 \\ 
350 & 1800 & 1.00419 & 0.99383 & 1.01309 & 1.02065 & 0.37842 & -0.00009 \\ 
350 & 5000 & 1.00176 & 0.99898 & 0.98630 & 0.97756 & 0.33294 & -0.00007 \\ 
500 & 200 & 1.01625 & 1.04360 & 0.95588 & 0.97053 & 0.28455 & -0.00446 \\ 
500 & 800 & 0.99254 & 1.01238 & 0.98620 & 0.96630 & 0.39177 & 0.00009 \\ 
500 & 1800 & 1.00419 & 0.99383 & 1.01309 & 1.02065 & 0.37842 & -0.00009 \\ 
500 & 5000 & 1.00176 & 0.99898 & 0.98630 & 0.97756 & 0.33294 & -0.00007 \\ 
650 & 200 & 1.01613 & 1.04374 & 0.94760 & 0.96836 & 0.28279 & -0.00445 \\ 
650 & 800 & 0.99254 & 1.01238 & 0.98620 & 0.96630 & 0.39177 & 0.00009 \\ 
650 & 1800 & 1.00419 & 0.99383 & 1.01309 & 1.02065 & 0.37842 & -0.00009 \\ 
650 & 5000 & 1.00176 & 0.99898 & 0.98630 & 0.97756 & 0.33294 & -0.00007 \\ 
800 & 200 & 1.01613 & 1.04374 & 0.94760 & 0.96836 & 0.28279 & -0.00445 \\ 
800 & 800 & 0.99254 & 1.01238 & 0.98620 & 0.96630 & 0.39177 & 0.00009 \\ 
800 & 1800 & 1.00419 & 0.99383 & 1.01309 & 1.02065 & 0.37842 & -0.00009 \\ 
800 & 5000 & 1.00176 & 0.99898 & 0.98630 & 0.97756 & 0.33294 & -0.00007 \\ 
\midrule\addlinespace[2.5pt]
\multicolumn{8}{l}{$\phi = 0.8$} \\ 
\midrule\addlinespace[2.5pt]
50 & 200 & 1.06940 & 1.03728 & 0.98856 & 0.98784 & 0.13615 & -0.00550 \\ 
50 & 800 & 0.99944 & 1.00970 & 1.02213 & 0.96549 & 0.33119 & -0.00004 \\ 
50 & 1800 & 1.00781 & 0.99269 & 0.99405 & 1.02971 & 0.29987 & -0.00009 \\ 
50 & 5000 & 1.00251 & 0.99892 & 0.98838 & 0.98174 & 0.26032 & -0.00007 \\ 
200 & 200 & 0.99335 & 1.03434 & 0.91985 & 0.97336 & 0.25843 & -0.00423 \\ 
200 & 800 & 0.99622 & 1.01254 & 0.97878 & 0.96251 & 0.32505 & 0.00010 \\ 
200 & 1800 & 1.00443 & 0.99383 & 1.00714 & 1.02065 & 0.31206 & -0.00009 \\ 
200 & 5000 & 1.00255 & 0.99898 & 0.98013 & 0.97756 & 0.26672 & -0.00007 \\ 
350 & 200 & 1.00173 & 1.03784 & 0.93345 & 0.97138 & 0.16260 & -0.00480 \\ 
350 & 800 & 0.99581 & 1.01238 & 0.97988 & 0.96630 & 0.32393 & 0.00009 \\ 
350 & 1800 & 1.00443 & 0.99383 & 1.00714 & 1.02065 & 0.31206 & -0.00009 \\ 
350 & 5000 & 1.00255 & 0.99898 & 0.98013 & 0.97756 & 0.26672 & -0.00007 \\ 
500 & 200 & 1.01486 & 1.04360 & 0.94984 & 0.97053 & 0.21650 & -0.00446 \\ 
500 & 800 & 0.99581 & 1.01238 & 0.97988 & 0.96630 & 0.32393 & 0.00009 \\ 
500 & 1800 & 1.00443 & 0.99383 & 1.00714 & 1.02065 & 0.31206 & -0.00009 \\ 
500 & 5000 & 1.00255 & 0.99898 & 0.98013 & 0.97756 & 0.26672 & -0.00007 \\ 
650 & 200 & 1.01471 & 1.04374 & 0.94179 & 0.96836 & 0.21440 & -0.00445 \\ 
650 & 800 & 0.99581 & 1.01238 & 0.97988 & 0.96630 & 0.32393 & 0.00009 \\ 
650 & 1800 & 1.00443 & 0.99383 & 1.00714 & 1.02065 & 0.31206 & -0.00009 \\ 
650 & 5000 & 1.00255 & 0.99898 & 0.98013 & 0.97756 & 0.26672 & -0.00007 \\ 
800 & 200 & 1.01471 & 1.04374 & 0.94179 & 0.96836 & 0.21440 & -0.00445 \\ 
800 & 800 & 0.99581 & 1.01238 & 0.97988 & 0.96630 & 0.32393 & 0.00009 \\ 
800 & 1800 & 1.00443 & 0.99383 & 1.00714 & 1.02065 & 0.31206 & -0.00009 \\ 
800 & 5000 & 1.00255 & 0.99898 & 0.98013 & 0.97756 & 0.26672 & -0.00007 \\ 
\bottomrule
\end{longtable}
\begin{minipage}{\linewidth}
\textsuperscript{\textit{1}}We still report $\hat{\rho}_{FL}$, see Equation \ref{eq:model-2} for completeness purposes even though it can not be compared with $\hat{\psi}_{PL}$\\
\end{minipage}

\begin{longtable}{cccccccc}
\caption{Relative bias ($\mathbf{R B}$) and mean squared error ($\mathbf{M S E}$), showing the influence of neighbourhood radius adjustment on the performance of spatial regression modelling and for different sample size. }\label{tab:mc-sim-met-buff} \\ 
\toprule
 &  & \multicolumn{3}{c}{$\mathbf{R B} (\textbf{PL}, \textbf{FL})$} & \multicolumn{3}{c}{$\mathbf{M S E} (\textbf{PL}, \textbf{FL})$} \\ 
\cmidrule(lr){3-5} \cmidrule(lr){6-8}
buffer & $n$ & $\hat{\beta}$ & $\hat{\sigma}$ & $\hat{\psi}$ & $\hat{\beta}$ & $\hat{\sigma}$ & $\hat{\psi}$\\ 
\midrule\addlinespace[2.5pt]
\multicolumn{8}{l}{$\phi = 1$} \\ 
\midrule\addlinespace[2.5pt]
50 & 200 & $\begin{array}{c} (0.06990, \\ 0.03728) \end{array}$ & $\begin{array}{c} (0.01216, \\ 0.01216) \end{array}$ & $\begin{array}{c} (0.78635, \\ 1.00550) \end{array}$ & $\begin{array}{c} (0.00489, \\ 0.00139) \end{array}$ & $\begin{array}{c} (0.00014, \\ 0.00015) \end{array}$ & $\begin{array}{c} (0.61835, \\ 1.01103) \end{array}$ \\ 
50 & 800 & $\begin{array}{c} (0.00188, \\ 0.00970) \end{array}$ & $\begin{array}{c} (0.03451, \\ 0.03451) \end{array}$ & $\begin{array}{c} (0.60483, \\ 1.00004) \end{array}$ & $\begin{array}{c} (0.00000, \\ 0.00009) \end{array}$ & $\begin{array}{c} (0.00091, \\ 0.00119) \end{array}$ & $\begin{array}{c} (0.36582, \\ 1.00009) \end{array}$ \\ 
50 & 1800 & $\begin{array}{c} (0.00633, \\ 0.00731) \end{array}$ & $\begin{array}{c} (0.02971, \\ 0.02971) \end{array}$ & $\begin{array}{c} (0.63483, \\ 1.00009) \end{array}$ & $\begin{array}{c} (0.00004, \\ 0.00005) \end{array}$ & $\begin{array}{c} (0.00000, \\ 0.00088) \end{array}$ & $\begin{array}{c} (0.40301, \\ 1.00019) \end{array}$ \\ 
50 & 5000 & $\begin{array}{c} (0.00238, \\ 0.00108) \end{array}$ & $\begin{array}{c} (0.01826, \\ 0.01826) \end{array}$ & $\begin{array}{c} (0.67328, \\ 1.00007) \end{array}$ & $\begin{array}{c} (0.00001, \\ 0.00000) \end{array}$ & $\begin{array}{c} (0.00004, \\ 0.00033) \end{array}$ & $\begin{array}{c} (0.45331, \\ 1.00014) \end{array}$ \\ 
200 & 200 & $\begin{array}{c} (0.00539, \\ 0.03434) \end{array}$ & $\begin{array}{c} (0.02664, \\ 0.02664) \end{array}$ & $\begin{array}{c} (0.67159, \\ 1.00423) \end{array}$ & $\begin{array}{c} (0.00003, \\ 0.00118) \end{array}$ & $\begin{array}{c} (0.00555, \\ 0.00071) \end{array}$ & $\begin{array}{c} (0.45104, \\ 1.00847) \end{array}$ \\ 
200 & 800 & $\begin{array}{c} (0.00675, \\ 0.01254) \end{array}$ & $\begin{array}{c} (0.03749, \\ 0.03749) \end{array}$ & $\begin{array}{c} (0.60687, \\ 0.99990) \end{array}$ & $\begin{array}{c} (0.00005, \\ 0.00016) \end{array}$ & $\begin{array}{c} (0.00023, \\ 0.00141) \end{array}$ & $\begin{array}{c} (0.36829, \\ 0.99980) \end{array}$ \\ 
200 & 1800 & $\begin{array}{c} (0.00419, \\ 0.00617) \end{array}$ & $\begin{array}{c} (0.02065, \\ 0.02065) \end{array}$ & $\begin{array}{c} (0.62158, \\ 1.00009) \end{array}$ & $\begin{array}{c} (0.00002, \\ 0.00004) \end{array}$ & $\begin{array}{c} (0.00017, \\ 0.00043) \end{array}$ & $\begin{array}{c} (0.38636, \\ 1.00018) \end{array}$ \\ 
200 & 5000 & $\begin{array}{c} (0.00176, \\ 0.00102) \end{array}$ & $\begin{array}{c} (0.02244, \\ 0.02244) \end{array}$ & $\begin{array}{c} (0.66706, \\ 1.00007) \end{array}$ & $\begin{array}{c} (0.00000, \\ 0.00000) \end{array}$ & $\begin{array}{c} (0.00019, \\ 0.00050) \end{array}$ & $\begin{array}{c} (0.44497, \\ 1.00014) \end{array}$ \\ 
350 & 200 & $\begin{array}{c} (0.00347, \\ 0.03784) \end{array}$ & $\begin{array}{c} (0.02862, \\ 0.02862) \end{array}$ & $\begin{array}{c} (0.77098, \\ 1.00480) \end{array}$ & $\begin{array}{c} (0.00001, \\ 0.00143) \end{array}$ & $\begin{array}{c} (0.00351, \\ 0.00082) \end{array}$ & $\begin{array}{c} (0.59442, \\ 1.00963) \end{array}$ \\ 
350 & 800 & $\begin{array}{c} (0.00746, \\ 0.01238) \end{array}$ & $\begin{array}{c} (0.03370, \\ 0.03370) \end{array}$ & $\begin{array}{c} (0.60823, \\ 0.99991) \end{array}$ & $\begin{array}{c} (0.00006, \\ 0.00015) \end{array}$ & $\begin{array}{c} (0.00019, \\ 0.00114) \end{array}$ & $\begin{array}{c} (0.36995, \\ 0.99981) \end{array}$ \\ 
350 & 1800 & $\begin{array}{c} (0.00419, \\ 0.00617) \end{array}$ & $\begin{array}{c} (0.02065, \\ 0.02065) \end{array}$ & $\begin{array}{c} (0.62158, \\ 1.00009) \end{array}$ & $\begin{array}{c} (0.00002, \\ 0.00004) \end{array}$ & $\begin{array}{c} (0.00017, \\ 0.00043) \end{array}$ & $\begin{array}{c} (0.38636, \\ 1.00018) \end{array}$ \\ 
350 & 5000 & $\begin{array}{c} (0.00176, \\ 0.00102) \end{array}$ & $\begin{array}{c} (0.02244, \\ 0.02244) \end{array}$ & $\begin{array}{c} (0.66706, \\ 1.00007) \end{array}$ & $\begin{array}{c} (0.00000, \\ 0.00000) \end{array}$ & $\begin{array}{c} (0.00019, \\ 0.00050) \end{array}$ & $\begin{array}{c} (0.44497, \\ 1.00014) \end{array}$ \\ 
500 & 200 & $\begin{array}{c} (0.01625, \\ 0.04360) \end{array}$ & $\begin{array}{c} (0.02947, \\ 0.02947) \end{array}$ & $\begin{array}{c} (0.71545, \\ 1.00446) \end{array}$ & $\begin{array}{c} (0.00026, \\ 0.00190) \end{array}$ & $\begin{array}{c} (0.00195, \\ 0.00087) \end{array}$ & $\begin{array}{c} (0.51187, \\ 1.00894) \end{array}$ \\ 
500 & 800 & $\begin{array}{c} (0.00746, \\ 0.01238) \end{array}$ & $\begin{array}{c} (0.03370, \\ 0.03370) \end{array}$ & $\begin{array}{c} (0.60823, \\ 0.99991) \end{array}$ & $\begin{array}{c} (0.00006, \\ 0.00015) \end{array}$ & $\begin{array}{c} (0.00019, \\ 0.00114) \end{array}$ & $\begin{array}{c} (0.36995, \\ 0.99981) \end{array}$ \\ 
500 & 1800 & $\begin{array}{c} (0.00419, \\ 0.00617) \end{array}$ & $\begin{array}{c} (0.02065, \\ 0.02065) \end{array}$ & $\begin{array}{c} (0.62158, \\ 1.00009) \end{array}$ & $\begin{array}{c} (0.00002, \\ 0.00004) \end{array}$ & $\begin{array}{c} (0.00017, \\ 0.00043) \end{array}$ & $\begin{array}{c} (0.38636, \\ 1.00018) \end{array}$ \\ 
500 & 5000 & $\begin{array}{c} (0.00176, \\ 0.00102) \end{array}$ & $\begin{array}{c} (0.02244, \\ 0.02244) \end{array}$ & $\begin{array}{c} (0.66706, \\ 1.00007) \end{array}$ & $\begin{array}{c} (0.00000, \\ 0.00000) \end{array}$ & $\begin{array}{c} (0.00019, \\ 0.00050) \end{array}$ & $\begin{array}{c} (0.44497, \\ 1.00014) \end{array}$ \\ 
650 & 200 & $\begin{array}{c} (0.01613, \\ 0.04374) \end{array}$ & $\begin{array}{c} (0.03164, \\ 0.03164) \end{array}$ & $\begin{array}{c} (0.71721, \\ 1.00445) \end{array}$ & $\begin{array}{c} (0.00026, \\ 0.00191) \end{array}$ & $\begin{array}{c} (0.00275, \\ 0.00100) \end{array}$ & $\begin{array}{c} (0.51439, \\ 1.00893) \end{array}$ \\ 
650 & 800 & $\begin{array}{c} (0.00746, \\ 0.01238) \end{array}$ & $\begin{array}{c} (0.03370, \\ 0.03370) \end{array}$ & $\begin{array}{c} (0.60823, \\ 0.99991) \end{array}$ & $\begin{array}{c} (0.00006, \\ 0.00015) \end{array}$ & $\begin{array}{c} (0.00019, \\ 0.00114) \end{array}$ & $\begin{array}{c} (0.36995, \\ 0.99981) \end{array}$ \\ 
650 & 1800 & $\begin{array}{c} (0.00419, \\ 0.00617) \end{array}$ & $\begin{array}{c} (0.02065, \\ 0.02065) \end{array}$ & $\begin{array}{c} (0.62158, \\ 1.00009) \end{array}$ & $\begin{array}{c} (0.00002, \\ 0.00004) \end{array}$ & $\begin{array}{c} (0.00017, \\ 0.00043) \end{array}$ & $\begin{array}{c} (0.38636, \\ 1.00018) \end{array}$ \\ 
650 & 5000 & $\begin{array}{c} (0.00176, \\ 0.00102) \end{array}$ & $\begin{array}{c} (0.02244, \\ 0.02244) \end{array}$ & $\begin{array}{c} (0.66706, \\ 1.00007) \end{array}$ & $\begin{array}{c} (0.00000, \\ 0.00000) \end{array}$ & $\begin{array}{c} (0.00019, \\ 0.00050) \end{array}$ & $\begin{array}{c} (0.44497, \\ 1.00014) \end{array}$ \\ 
800 & 200 & $\begin{array}{c} (0.01471, \\ 0.04374) \end{array}$ & $\begin{array}{c} (0.03164, \\ 0.03164) \end{array}$ & $\begin{array}{c} (0.98199, \\ 1.25557) \end{array}$ & $\begin{array}{c} (0.00022, \\ 0.00191) \end{array}$ & $\begin{array}{c} (0.00339, \\ 0.00100) \end{array}$ & $\begin{array}{c} (0.61716, \\ 1.00893) \end{array}$ \\ 
800 & 800 & $\begin{array}{c} (0.00419, \\ 0.01238) \end{array}$ & $\begin{array}{c} (0.03370, \\ 0.03370) \end{array}$ & $\begin{array}{c} (0.84509, \\ 1.24988) \end{array}$ & $\begin{array}{c} (0.00002, \\ 0.00015) \end{array}$ & $\begin{array}{c} (0.00040, \\ 0.00114) \end{array}$ & $\begin{array}{c} (0.45708, \\ 0.99981) \end{array}$ \\ 
800 & 1800 & $\begin{array}{c} (0.00443, \\ 0.00617) \end{array}$ & $\begin{array}{c} (0.02065, \\ 0.02065) \end{array}$ & $\begin{array}{c} (0.85993, \\ 1.25011) \end{array}$ & $\begin{array}{c} (0.00002, \\ 0.00004) \end{array}$ & $\begin{array}{c} (0.00005, \\ 0.00043) \end{array}$ & $\begin{array}{c} (0.47326, \\ 1.00018) \end{array}$ \\ 
800 & 5000 & $\begin{array}{c} (0.00255, \\ 0.00102) \end{array}$ & $\begin{array}{c} (0.02244, \\ 0.02244) \end{array}$ & $\begin{array}{c} (0.91661, \\ 1.25009) \end{array}$ & $\begin{array}{c} (0.00001, \\ 0.00000) \end{array}$ & $\begin{array}{c} (0.00039, \\ 0.00050) \end{array}$ & $\begin{array}{c} (0.53771, \\ 1.00014) \end{array}$ \\ 
\midrule\addlinespace[2.5pt]
\multicolumn{8}{l}{$\phi = 0.8$} \\ 
\midrule\addlinespace[2.5pt]
50 & 200 & $\begin{array}{c} (0.06940, \\ 0.03728) \end{array}$ & $\begin{array}{c} (0.01216, \\ 0.01216) \end{array}$ & $\begin{array}{c} (1.07981, \\ 1.25687) \end{array}$ & $\begin{array}{c} (0.00482, \\ 0.00139) \end{array}$ & $\begin{array}{c} (0.00013, \\ 0.00015) \end{array}$ & $\begin{array}{c} (0.74624, \\ 1.01103) \end{array}$ \\ 
50 & 800 & $\begin{array}{c} (0.00056, \\ 0.00970) \end{array}$ & $\begin{array}{c} (0.03451, \\ 0.03451) \end{array}$ & $\begin{array}{c} (0.83601, \\ 1.25005) \end{array}$ & $\begin{array}{c} (0.00000, \\ 0.00009) \end{array}$ & $\begin{array}{c} (0.00049, \\ 0.00119) \end{array}$ & $\begin{array}{c} (0.44731, \\ 1.00009) \end{array}$ \\ 
50 & 1800 & $\begin{array}{c} (0.00781, \\ 0.00731) \end{array}$ & $\begin{array}{c} (0.02971, \\ 0.02971) \end{array}$ & $\begin{array}{c} (0.87516, \\ 1.25012) \end{array}$ & $\begin{array}{c} (0.00006, \\ 0.00005) \end{array}$ & $\begin{array}{c} (0.00004, \\ 0.00088) \end{array}$ & $\begin{array}{c} (0.49018, \\ 1.00019) \end{array}$ \\ 
50 & 5000 & $\begin{array}{c} (0.00251, \\ 0.00108) \end{array}$ & $\begin{array}{c} (0.01826, \\ 0.01826) \end{array}$ & $\begin{array}{c} (0.92460, \\ 1.25009) \end{array}$ & $\begin{array}{c} (0.00001, \\ 0.00000) \end{array}$ & $\begin{array}{c} (0.00013, \\ 0.00033) \end{array}$ & $\begin{array}{c} (0.54713, \\ 1.00014) \end{array}$ \\ 
200 & 200 & $\begin{array}{c} (0.00665, \\ 0.03434) \end{array}$ & $\begin{array}{c} (0.02664, \\ 0.02664) \end{array}$ & $\begin{array}{c} (0.92696, \\ 1.25529) \end{array}$ & $\begin{array}{c} (0.00004, \\ 0.00118) \end{array}$ & $\begin{array}{c} (0.00642, \\ 0.00071) \end{array}$ & $\begin{array}{c} (0.54992, \\ 1.00847) \end{array}$ \\ 
200 & 800 & $\begin{array}{c} (0.00378, \\ 0.01254) \end{array}$ & $\begin{array}{c} (0.03749, \\ 0.03749) \end{array}$ & $\begin{array}{c} (0.84369, \\ 1.24988) \end{array}$ & $\begin{array}{c} (0.00001, \\ 0.00016) \end{array}$ & $\begin{array}{c} (0.00045, \\ 0.00141) \end{array}$ & $\begin{array}{c} (0.45556, \\ 0.99980) \end{array}$ \\ 
200 & 1800 & $\begin{array}{c} (0.00443, \\ 0.00617) \end{array}$ & $\begin{array}{c} (0.02065, \\ 0.02065) \end{array}$ & $\begin{array}{c} (0.85993, \\ 1.25011) \end{array}$ & $\begin{array}{c} (0.00002, \\ 0.00004) \end{array}$ & $\begin{array}{c} (0.00005, \\ 0.00043) \end{array}$ & $\begin{array}{c} (0.47326, \\ 1.00018) \end{array}$ \\ 
200 & 5000 & $\begin{array}{c} (0.00255, \\ 0.00102) \end{array}$ & $\begin{array}{c} (0.02244, \\ 0.02244) \end{array}$ & $\begin{array}{c} (0.91661, \\ 1.25009) \end{array}$ & $\begin{array}{c} (0.00001, \\ 0.00000) \end{array}$ & $\begin{array}{c} (0.00039, \\ 0.00050) \end{array}$ & $\begin{array}{c} (0.53771, \\ 1.00014) \end{array}$ \\ 
350 & 200 & $\begin{array}{c} (0.00173, \\ 0.03784) \end{array}$ & $\begin{array}{c} (0.02862, \\ 0.02862) \end{array}$ & $\begin{array}{c} (1.04675, \\ 1.25600) \end{array}$ & $\begin{array}{c} (0.00000, \\ 0.00143) \end{array}$ & $\begin{array}{c} (0.00443, \\ 0.00082) \end{array}$ & $\begin{array}{c} (0.70123, \\ 1.00963) \end{array}$ \\ 
350 & 800 & $\begin{array}{c} (0.00419, \\ 0.01238) \end{array}$ & $\begin{array}{c} (0.03370, \\ 0.03370) \end{array}$ & $\begin{array}{c} (0.84509, \\ 1.24988) \end{array}$ & $\begin{array}{c} (0.00002, \\ 0.00015) \end{array}$ & $\begin{array}{c} (0.00040, \\ 0.00114) \end{array}$ & $\begin{array}{c} (0.45708, \\ 0.99981) \end{array}$ \\ 
350 & 1800 & $\begin{array}{c} (0.00443, \\ 0.00617) \end{array}$ & $\begin{array}{c} (0.02065, \\ 0.02065) \end{array}$ & $\begin{array}{c} (0.85993, \\ 1.25011) \end{array}$ & $\begin{array}{c} (0.00002, \\ 0.00004) \end{array}$ & $\begin{array}{c} (0.00005, \\ 0.00043) \end{array}$ & $\begin{array}{c} (0.47326, \\ 1.00018) \end{array}$ \\ 
350 & 5000 & $\begin{array}{c} (0.00255, \\ 0.00102) \end{array}$ & $\begin{array}{c} (0.02244, \\ 0.02244) \end{array}$ & $\begin{array}{c} (0.91661, \\ 1.25009) \end{array}$ & $\begin{array}{c} (0.00001, \\ 0.00000) \end{array}$ & $\begin{array}{c} (0.00039, \\ 0.00050) \end{array}$ & $\begin{array}{c} (0.53771, \\ 1.00014) \end{array}$ \\ 
500 & 200 & $\begin{array}{c} (0.01486, \\ 0.04360) \end{array}$ & $\begin{array}{c} (0.02947, \\ 0.02947) \end{array}$ & $\begin{array}{c} (0.97937, \\ 1.25558) \end{array}$ & $\begin{array}{c} (0.00022, \\ 0.00190) \end{array}$ & $\begin{array}{c} (0.00252, \\ 0.00087) \end{array}$ & $\begin{array}{c} (0.61387, \\ 1.00894) \end{array}$ \\ 
500 & 800 & $\begin{array}{c} (0.00419, \\ 0.01238) \end{array}$ & $\begin{array}{c} (0.03370, \\ 0.03370) \end{array}$ & $\begin{array}{c} (0.84509, \\ 1.24988) \end{array}$ & $\begin{array}{c} (0.00002, \\ 0.00015) \end{array}$ & $\begin{array}{c} (0.00040, \\ 0.00114) \end{array}$ & $\begin{array}{c} (0.45708, \\ 0.99981) \end{array}$ \\ 
500 & 1800 & $\begin{array}{c} (0.00443, \\ 0.00617) \end{array}$ & $\begin{array}{c} (0.02065, \\ 0.02065) \end{array}$ & $\begin{array}{c} (0.85993, \\ 1.25011) \end{array}$ & $\begin{array}{c} (0.00002, \\ 0.00004) \end{array}$ & $\begin{array}{c} (0.00005, \\ 0.00043) \end{array}$ & $\begin{array}{c} (0.47326, \\ 1.00018) \end{array}$ \\ 
500 & 5000 & $\begin{array}{c} (0.00255, \\ 0.00102) \end{array}$ & $\begin{array}{c} (0.02244, \\ 0.02244) \end{array}$ & $\begin{array}{c} (0.91661, \\ 1.25009) \end{array}$ & $\begin{array}{c} (0.00001, \\ 0.00000) \end{array}$ & $\begin{array}{c} (0.00039, \\ 0.00050) \end{array}$ & $\begin{array}{c} (0.53771, \\ 1.00014) \end{array}$ \\ 
650 & 200 & $\begin{array}{c} (0.01471, \\ 0.04374) \end{array}$ & $\begin{array}{c} (0.03164, \\ 0.03164) \end{array}$ & $\begin{array}{c} (0.98199, \\ 1.25557) \end{array}$ & $\begin{array}{c} (0.00022, \\ 0.00191) \end{array}$ & $\begin{array}{c} (0.00339, \\ 0.00100) \end{array}$ & $\begin{array}{c} (0.61716, \\ 1.00893) \end{array}$ \\ 
650 & 800 & $\begin{array}{c} (0.00419, \\ 0.01238) \end{array}$ & $\begin{array}{c} (0.03370, \\ 0.03370) \end{array}$ & $\begin{array}{c} (0.84509, \\ 1.24988) \end{array}$ & $\begin{array}{c} (0.00002, \\ 0.00015) \end{array}$ & $\begin{array}{c} (0.00040, \\ 0.00114) \end{array}$ & $\begin{array}{c} (0.45708, \\ 0.99981) \end{array}$ \\ 
650 & 1800 & $\begin{array}{c} (0.00443, \\ 0.00617) \end{array}$ & $\begin{array}{c} (0.02065, \\ 0.02065) \end{array}$ & $\begin{array}{c} (0.85993, \\ 1.25011) \end{array}$ & $\begin{array}{c} (0.00002, \\ 0.00004) \end{array}$ & $\begin{array}{c} (0.00005, \\ 0.00043) \end{array}$ & $\begin{array}{c} (0.47326, \\ 1.00018) \end{array}$ \\ 
650 & 5000 & $\begin{array}{c} (0.00255, \\ 0.00102) \end{array}$ & $\begin{array}{c} (0.02244, \\ 0.02244) \end{array}$ & $\begin{array}{c} (0.91661, \\ 1.25009) \end{array}$ & $\begin{array}{c} (0.00001, \\ 0.00000) \end{array}$ & $\begin{array}{c} (0.00039, \\ 0.00050) \end{array}$ & $\begin{array}{c} (0.53771, \\ 1.00014) \end{array}$ \\ 
800 & 200 & $\begin{array}{c} (0.01471, \\ 0.04374) \end{array}$ & $\begin{array}{c} (0.03164, \\ 0.03164) \end{array}$ & $\begin{array}{c} (0.98199, \\ 1.25557) \end{array}$ & $\begin{array}{c} (0.00022, \\ 0.00191) \end{array}$ & $\begin{array}{c} (0.00339, \\ 0.00100) \end{array}$ & $\begin{array}{c} (0.61716, \\ 1.00893) \end{array}$ \\ 
800 & 800 & $\begin{array}{c} (0.00419, \\ 0.01238) \end{array}$ & $\begin{array}{c} (0.03370, \\ 0.03370) \end{array}$ & $\begin{array}{c} (0.84509, \\ 1.24988) \end{array}$ & $\begin{array}{c} (0.00002, \\ 0.00015) \end{array}$ & $\begin{array}{c} (0.00040, \\ 0.00114) \end{array}$ & $\begin{array}{c} (0.45708, \\ 0.99981) \end{array}$ \\ 
800 & 1800 & $\begin{array}{c} (0.00443, \\ 0.00617) \end{array}$ & $\begin{array}{c} (0.02065, \\ 0.02065) \end{array}$ & $\begin{array}{c} (0.85993, \\ 1.25011) \end{array}$ & $\begin{array}{c} (0.00002, \\ 0.00004) \end{array}$ & $\begin{array}{c} (0.00005, \\ 0.00043) \end{array}$ & $\begin{array}{c} (0.47326, \\ 1.00018) \end{array}$ \\ 
800 & 5000 & $\begin{array}{c} (0.00255, \\ 0.00102) \end{array}$ & $\begin{array}{c} (0.02244, \\ 0.02244) \end{array}$ & $\begin{array}{c} (0.91661, \\ 1.25009) \end{array}$ & $\begin{array}{c} (0.00001, \\ 0.00000) \end{array}$ & $\begin{array}{c} (0.00039, \\ 0.00050) \end{array}$ & $\begin{array}{c} (0.53771, \\ 1.00014) \end{array}$ \\ 
\bottomrule
\end{longtable}

\section{Summary and conclusions} \label{concl}

The results obtained in this paper show that the KD-T PL algorithm not only provides accurate parameter estimates, but also offers significant advantages in terms of computational efficiency when compared to the full Maximum Likelihood (PL) approach.

Our findings, reveal that both PL and FL methods exhibit low relative bias and consistently low MSE further showing that the advantages of PL estimation extend beyond its computational efficiency.

Of particular note is the substantial computational efficiency demonstrated by the coupled KD-T PL algorithm, which significantly reduces the computational time required with respect to a full ML approach. This efficiency renders PL estimation particularly appealing for applications involving large spatial datasets, where time constraints are a critical concern. 

Further exploration of PL estimation in real-world applications is necessary to fully evaluate its benefits. Furthermore, it is worth noting that the PL method's versatility may extend beyond the spatial model considered in the present study, including alternatives such as non linear or spatial panel data models. Testing the PL estimation method on different spatial models as in \cite{wang_2013_partial} is an intriguing path for future research.

\bibliography{pairwise}

\end{document}